\documentclass{article}

\usepackage{arxiv}

\usepackage[utf8]{inputenc} 
\usepackage[T1]{fontenc}    
\usepackage{hyperref}       
\usepackage{url}            
\usepackage{booktabs}       
\usepackage{amsfonts}       
\usepackage{nicefrac}       
\usepackage{microtype}      
\usepackage{lipsum}
\usepackage{graphicx}
\usepackage{subfig}
\graphicspath{ {./images/} }

\title{Internal structure and heat conduction in rigid solids: a two temperature approach}

\author{
 R.E Gonzalez-Narvaez \\
   Centro de Investigación en Ciencias (CInC)\\
  Universidad Autónoma del Estado de Morelos (UAEM)\\
  Col. Chamilpa, Cuernavaca, México \\
  \texttt{ruth.gonzalez@uaem.edu.mx} \\
   \And
M. L\'opez de Haro \\
   Instituto de Energías Renovables\\
  Universidad Nacional Autónoma de México (U.N.A.M.)\\
  Temixco, Morelos, México \\
  \texttt{malopez@unam.mx} \\
  \And
  F. V\'azquez \\
  Centro de Investigación en Ciencias (CInC)\\
  Universidad Autónoma del Estado de Morelos (UAEM)\\
  Col. Chamilpa, Cuernavaca, México \\
  \texttt{vazquez@uaem.mx} \\
}

\begin{document}
\maketitle
\begin{abstract}
A non-Fourier thermal transport regime characterizes the heat conduction in solids with internal structure. Several thermodynamic theories attempt to explain the separation from the Fourier regime in such kind of systems. Here we develop a two temperature model to describe the non-Fourier regime from the principles of non-equilibrium thermodynamics. The basic assumption is the existence of two well separated length scales in the system, namely, one related with the matrix dimension (bulk) and the other with the characteristic length of the internal structure. Two Fourier type coupled transport equations are obtained for the temperatures which describe the heat conduction in each of the length scales. Recent experimental results from several groups on the thermal response of different structured materials are satisfactorily reproduced by using the coupling parameter as a fitting parameter. The similarities and differences of the present formalism with other theories are discussed. 
\end{abstract}


\section{Introduction}
Real systems and materials are non-homogeneous either concerning their transport properties or their inner structure (here onwards called the microstructure). Heterogeneity may arise from compactation, packing or sintering processes \cite{lee2002, huang2003,li2006}. The effects of inhomogeneity emerge from smaller length scales and they influence the response of the system to external thermodynamic forces. Irreversible thermodynamic formalisms have been formulated to include the microstructure in extended thermodynamic variable spaces through the so called internal variables \cite{mauguin1990,muschik1994a,muschik1994b,van2003,mauguin2015} and these formalisms have been applied to the case of a rigid heat conductor \cite{van2008,van2012,berezovski2012,berezovski2016,berezovski2017,berezovski2018}. For a survey of published papers on this topic before $1990$, see \cite{tamma1998}. Theoretically and experimentally it has been found that energy transport in solids with microstructure is dominated by non-Fourier mechanisms of heat transport, in such a way that in order to account for these  mechanisms the use of non-Fourier models like the Guyer-Krumhansl (GK) equation and others have been required.  One may reasonably wonder whether the Fourier transport equation is able to reproduce the non-Fourier behavior of microstructured solids by introducing somehow the structure features in the description. If one attempts to introduce the structure by making the heat conductivity to depend on the position, the answer is no as it will be shown below.
In this work, however,  we argue that it is possible to reproduce the non-Fourier behavior of microstrured solids if one assumes that there exist two basic heat transport processes which occur at different length scales: on the one hand,  heat transport in a macroscopic length scale comparable with the characteristic length of the macroscopic system; on the other,  transport processes occurring at a microscopic scale of length determined by the characteristic length of the microstructure.  This readily leads to the consideration of two well differentiated length scales in the system determined,  respectively,  by the characteristic length of the microstructure and by the macroscopic dimensions of the system itself.  We focus on elucidating if a two temperature model,  based on  Fourier transport equations for each of the above mentioned processes,  is suitable to describe the assumed macro and micro-processes and to explain the behaviour of the measurable  temperature.  The two temperature model framework is based on the fact that the total internal energy of the system may be separated in two parts, namely, the part associated with the macroscopic length scale processes and the part related to processes taking place in the microscopic length scale. In our approach, the local equilibrium hypothesis is taken into account by assuming that the local entropy depends on the internal energy of both macro and micro-processes.  In turn,  as it is usually done in two temperature modeling, each of the internal energies  is related to a quasi-temperature referred to as the macro and the micro-temperature, respectively.  The total measurable temperature of the system is then obtained through a weighted addition of the macroscopic and the microscopic quasi-temperatures. Thus, the measurable local temperature will be understood as an average temperature in the control volume as it is done in classical irreversible thermodynamics. Mention must be made of the fact that the above statements evokes that the heat carrier population is being separated into two parts as it is done in many heat transport problems. For instance, this is done in transient ballistic-diffusive heat transport where the total internal energy per unit volume is the sum of longitudinal and transverse phononic branches \cite{ma2013,wang2019}. Also, a similar separation is done in heat transport in common materials at the nanoscale, where the electrons and the phonons that transport energy are out of an equilibrium state during the heat transfer \cite{selliro2020,rogolino2017,sobolev2016,chen2002,kovacs2018,vazquez2020}, or in nonlocal thermoelectric transport in thin layers \cite{jou2014}. In any case, however, this point will not deserve further consideration in this work. Instead, we will stay here within the scope of phenomenological irreversible thermodynamics.

Our development is as follows. The time evolution equations for each one of the temperatures will be obtained from the local equilibrium hypothesis, as stated above, through the use of the conservation equations for the corresponding internal energies at each length scale.  Such equations lead directly to the well known entropy balance equation from which one may identify the "thermodynamic forces" conjugate to the respective fluxes by examining the entropy production term \cite{garciacolin2009}. We will show below that including the micro-processes leads to the non-Fourier character of the macroprocess. The microstructure will be incorporated into the scheme through small length scale inhomogeneities in the spatial distribution of the thermal conductivity of the material. Hence, the form of the thermal conductivity will in turn  characterize the microstrure.  Comparison with recent  theoretical and experimental results obtained in structured solids at room temperature will provide support to the proposed model.  We will compare our scheme, in particular, with the thermodynamic theory of internal variables \cite{berezovski2018}. An interesting aspect to be considered is the effect of the microstructure on the entropy production in the material, since the microstructure introduces an additional dissipation mechanism (some authors call it the excess entropy production \cite{bedeauxlibro}). Once this is done, it is then natural to explore its relation with extremum principles.

We close this section by summarizing our purposes in this paper. The first one is, using irreversible thermodynamics, to formulate a two temperature model to describe heat transport in a rigid heat conductor with internal structure at room temperature.  Our aim is to have a physical interpretation of the two introduced quasi-temperatures in terms of processes occurring at macroscopic length scales (of the order of magnitude of the system's size) and at microscopic length scales (of the order of magnitude of the size of the system's microstructure). Secondly, we want to compare our theoretical results with recent experimental reports on non-Fourier heat transport found in structured solids at room temperature and to analyze cases where the Onsager reciprocity relations are satisfied and cases where they are not. Finally,  we want to discuss the relation of the present formalism with other thermodynamic theories used to describe structured solid heat conductors.  To this end, the paper is organized as follows. In the next section we introduce the physical model and proceed with the irreversible thermodynamic formulation of heat transport in such a system. This is followed by sec.  \ref{sec3} in which the time evolution of the measurable temperature and its relation to the microstructure is addressed.  In both previous sections near equilibrium and far from equilibrium situations are considered and a comparison with experimental results in various systems is performed. Section \ref{sec4} deals with the relation between the entropy production and the microstructure. The paper is closed in the final section with further discussion and some concluding remarks.

\section{The physical model}
\label{sec2}

Our model considers the existence of two separate spatial scales in the system. On the one hand, the scale of the bulk of the material and on the other, the scale of the microstructure. So we consider that  there are two interacting subsystems involved in the transport processes at the two spatial scales. This implies that there are three sources of dissipation: i) processes on the macroscopic spatial scale, ii) processes at the microscopic scale and iii)  interaction processes - energy exchange - between the two subsystems. To each of the subsystems will correspond a part of the total internal energy of the system determined by the caloric properties of each one.  In agreement with the local equilibrium hypothesis, the thermodynamic space of states will be made up of both internal energies to which two quasi-temperatures will be associated. The total temperature will be determined by the caloric properties of the subsystems.It is worth mentioning that the two subsystems are analogous to what in other two-temperature models is called a mixture of gases made up of different heat carriers, eg, electrons and phonons in the case of thermoelectricity. In Ref.  \cite{sellito2016}, for example, a “logic” argument is used that leads to the assumption that the respective energies of each gas are part of the space of state variables. We do not resort to such kind of argument but instead let the comparison with experimental results indicate whether the
assumption was adequate.

\subsection{Thermodynamic formulation of the time evolution equations}
\label{sec:thermo}
As already mentioned, the system consists of a rigid heat conducting solid with an internal structure in which two coupled irreversible processes take place: on the one hand, processes that occur at a macroscopic length scale, that is at a length scale of the same order of magnitude as the characteristic length of the system itself $L$; on the other hand,  processes that occur at a length scale of the same order of magnitude as the characteristic length of the internal structure $l$. The starting assumption is that the total internal energy of the system can be divided into two parts, each one corresponding to the above two dissipative processes, namely
\begin{equation}
 u=u_{M}+u_{m},
 \label{eq:interna}
\end{equation}
where $u$ stands for the internal energy and the labels $M$ and $m$ denote macroscopic and microscopic processes, respectively. 
From now on, we will assume that $L \gg l$ and we will refer to the microstructure as the internal structure of the system. Defining two quasi-temperatures through  the known caloric relation, $u=CT$ with $C$ the heat capacity at constant volume, we have the following
\begin{equation}
 T_{M}=\frac{u_{M}}{C_{M}},
 \label{eq:cuasimacro}
\end{equation}
and
\begin{equation}
 T_{m}=\frac{u_{m}}{C_{m}}.
 \label{eq:cuasimicro}
\end{equation}
Thus, the total (measurable) temperature of the system is given by the average
\begin{equation}
 T=\frac{C_{M}T_{M}+C_{m}T_{m}}{C_{M}+C_{m}}.
 \label{eq:total}
\end{equation}
The temperatures $T_{M}$ and $T_{m}$ will be referred to as  macro-temperature and micro-temperature, respectively. Note that, for the sake of brevity, we have omitted the prefix quasi in both instances. The macro-temperature is an average temperature that describes the processes that occur in the bulk and the micro-temperature describes the changes that frequently and slightly separate the system from the average temporal evolution due to the emergence of effects originating from the microstructure. Strictly speaking, and given that $L \gg l$, the micro-processes take place in a medium with  heterogeneous thermal conductivity. Meanwhile, macro-processes occur in an effective homogeneous environment. As can be seen, this approach conforms a two-temperature model for heat transport in the rigid solid where one of the temperatures accounts for the deviation of the system from the Fourier behavior. 

We next obtain the expression for the entropy production in the solid. We proceed in the usual way as it is done in two temperature models within the framework of irreversible thermodynamics \cite{sellito2016}. The starting assumption is that, as mentioned above,  the thermodynamic space of variables is constituted by the internal energies $u_{M}$ and $u_{m}$ in such a way that the entropy is a function of them,  namely
\begin{equation}
 S=S\left(u_{M},u_{m}\right).
 \label{eq:entropia}
\end{equation}
The time evolution of the total entropy of the system is then
\begin{equation}
 \frac{dS}{dt}=\left(\frac{\partial S}{\partial u_{M}}\right)_{u_{m}}\frac{du_{M}}{dt}+\left(\frac{\partial S}{\partial u_{m}}\right)_{u_{M}}\frac{du_{m}}{dt},
 \label{eq:entropiaevolution}
\end{equation}
where
\begin{equation}
 \left(\frac{\partial S}{\partial u_{M}}\right)_{u_{m}}=\frac{1}{T_{M}},
  \label{eq:macrotemp}
\end{equation}
and
\begin{equation}
 \left(\frac{\partial S}{\partial u_{m}}\right)_{u_{M}}=\frac{1}{T_{m}}.
  \label{eq:microtemp}
\end{equation}
Additionally, we will assume that both types of processes in the solid are conservative, that is, we will ignore the existence of energy sources.  Under those conditions
\begin{equation}
 \frac{du_{M}}{dt}=-\frac{\partial q_{M}}{\partial x},
  \label{eq:macrocons}
\end{equation}
and
\begin{equation}
 \frac{du_{m}}{dt}=-\frac{\partial q_{m}}{\partial x}.
  \label{eq:microcons}
\end{equation}
For simplicity, here we have constrained the system to one dimension and $q_{M}$ and $q_{m}$ are heat fluxes associated with energy transport in macro and micro-processes, respectively.
From Eqs. (\ref{eq:entropiaevolution})-(\ref{eq:microcons}) it is straightforward to get the balance equation of the total entropy, which reads 
\begin{equation}
 \frac{dS}{dt}+\frac{\partial}{\partial x}\left(\frac{1}{T_{M}}q_{M}+\frac{1}{T_{m}}q_{m}\right)=-\frac{1}{T_{M}}\left(q_{M}\frac{\frac{\partial T_{M}}{\partial x}}{T_{M}}\right)-\frac{1}{T_{m}}\left(q_{m}\frac{\frac{\partial T_{m}}{\partial x}}{T_{m}}\right).
 \label{eq:entropiabalance}
\end{equation}
In this last equation the right hand side is the entropy production $\sigma$,  \emph{i.e.}
\begin{equation}
 \sigma=-\frac{1}{T_{M}}\left(q_{M}\frac{\frac{\partial T_{M}}{\partial x}}{T_{M}}\right)-\frac{1}{T_{m}}\left(q_{m}\frac{\frac{\partial T_{m}}{\partial x}}{T_{m}}\right).
 \label{eq:entropyprod}
\end{equation}
In order to have a closed system of equations,  one has to relate the fluxes with the thermodynamic forces.  Note that in this case the fluxes and forces have the same tensorial character. In the absence of any further physical insight, the simplest way to relate such forces and fluxes is through linear relationships.  Hence, we will assume that
\begin{equation}
 -q_{M}=L_{11}\frac{\frac{\partial T_{M}}{\partial x}}{T_{M}^{2}}+L_{12}\frac{\frac{\partial T_{m}}{\partial x}}{T_{m}^{2}},
 \label{eq:constitutive1}
\end{equation}
\begin{equation}
 -q_{m}=L_{21}\frac{\frac{\partial T_{M}}{\partial x}}{T_{M}^{2}}+L_{22}\frac{\frac{\partial T_{m}}{\partial x}}{T_{m}^{2}},
 \label{eq:constitutive2}
\end{equation}
where the coefficients $L_{ij}$ are phenomenological coefficients with $i,j \epsilon \left[[1,2\right]]$. It is worth pointing out that they may be functions of the temperatures $T_{M}$ and $T_{m}$.  Further, depending on the actual physical situation,  such coefficients may or may not satisfy the Onsager reciprocity relations. Both cases will be considered. Equations (\ref{eq:constitutive1}) and (\ref{eq:constitutive2}) are the simplest way to ensure the non-negativity of entropy production. The proposed formalism is compatible with the second law if the Onsager coefficients matrix is positive definite. The validity of the Onsager relations only implies that this matrix is symmetric. So, we have expressed positiveness of the matrix as $L_{11}> 0$, $L_{22}> 0$, $L_{11}L_{22}-\left( L_{12} + L_{21}\right)^2/4> 0$, and we have ensured that these conditions are satisfied in all the studied examples.

\subsection{Near equilibrium processes. Validity of Onsager reciprocity relations.}
\label{sec:onsager}
We begin with the case of a system undergoing a slow process where it is allowed to reach the local equilibrium state at each stage. Then, we assume that the Onsager reciprocity relations are valid, \emph{ i.e.}
\begin{equation}
L_{12}=L_{21},
\end{equation}
which implies that 
\begin{equation} 
\lambda_{Mm}T_{m}^{2}=\lambda_{mM}T_{M}^{2}.
\label{eq:ORs}
\end{equation}

One of the necessary assumptions for the validity of the Onsager relations, ORs, is that the state variables are well defined and independent. According to Irreversible Thermodynamics principles, state variables satisfy both requirements in local equilibrium states. Thus, the local equilibrium hypothesis is a necessary condition for the validity of the Onsager relations. The other conditions of the validity of the ORs are also satisfied by the method used, that is, the definition of the fluxes and conjugated forces from the second law, and a linear relationship between both groups of variables. In the case
in which the system does not transit through local equilibrium states, as it is in the first example considered in the following section, the first of the mentioned conditions is not ensured.
In this case, the reproduction of the experimental results as observed in the respective figures, will give the assumed hypothesis -that is, the validity of the Onsager relations- an acceptable degree of plausibility.

Nonlinear transport equations for the temperatures are then obtained which read

\begin{eqnarray}
 C_{M}T_{m}^{3}\frac{dT_{M}}{dt}=\lambda_{M}T_{m}^{3}\frac{\partial^{2} T_{M}}{\partial x^{2}}+\lambda_{mM}\left[T_{M}^{2}T_{m}\frac{\partial^{2} T_{m}}{\partial x^{2}}-2T_{M}^{2}
 \left(\frac{\partial T_{m}}{\partial x}\right)^{2}\right]
 +\nonumber\\
2 \lambda_{mM}T_{M}T_{m}\frac{\partial T_{M}}{\partial x}\frac{\partial T_{m}}{\partial x},
 \label{eq:nonltransportM}
\end{eqnarray}

\begin{equation}
 C_{m}\frac{dT_{m}}{dt}=\lambda_{mM}\frac{\partial^{2} T_{M}}{\partial x^{2}}+\lambda_{m}\frac{\partial^{2} T_{m}}{\partial x^{2}}+\frac{\partial \lambda_{m}}{\partial x}\frac{\partial T_{m}}{\partial x}.
 \label{eq:nonltransportm}
\end{equation}

Note that by writing the constitutive relations as in Eqs. (\ref{eq:constitutive1}) and (\ref{eq:constitutive2}), we are implying that the time evolution of the system is described by two coupled Fourier,  though nonlinear,  type  processes.
It is worth remarking that, in this case, the assumption of the Onsager reciprocity relations leads to nonlinear transport equations.

As mentioned earlier, the material microstructure will be introduced in the two temperature model by assuming that the thermal conductivity of micro-processes, $\lambda_{m}$, is a function of position.  In the following section we will study its effects on the profile and time evolution of the total temperature.  Furthermore, we will  also analyze these effects on the stationary global entropy production. 

\subsection{Far from equilibrium processes. Non validity of the Onsager reciprocity relations.}
\label{sec:nononsager}
In this section we deal with the case in which the Onsager reciprocity relations are not satisfied in Eqs. (\ref{eq:constitutive1}) and (\ref{eq:constitutive2}). This occurs when, for instance, the system is thermally perturbed in a very short time scale (short with respect its  characteristic time scale) as it is the case of heat pulsed rigid  conductors in which the system is suddenly and abruptly brought out of equilibrium. The resulting heat transport regime is of the non-Fourier type since thermal inertia influences its response.  It must then be clear that although the Onsager relations are not satisfied it is not unreasonable to expect that linear constitutive equations will suffice to describe suitably the heat transport in the structured solid.  Note that this situation is different from that where non-linearity determines the behavior of the system and may lead to non validity of the  Onsager relations \cite{rogolino2015}.

The first experiments on pulsed materials were performed in the 70's in cold crystals and recently they were replicated in some structured solids at room temperature \cite{tang2007,van2017,kovacs2018b}.
We apply Eqs. (\ref{eq:constitutive1}) and (\ref{eq:constitutive2}) to describe the thermal response of a material at room temperature. The phenomenological coefficients dependence on the temperatures is assumed to be $L_{11}=\lambda_{M}T_{M}^{2}$, $L_{12}=\lambda_{Mm}T_{m}^{2}$, $L_{21}=\lambda_{mM}T_{M}^{2}$,  $L_{22}=\lambda_{m}T_{m}^{2}$. Then by combining Eqs. (\ref{eq:constitutive1}) and (\ref{eq:constitutive2}) with the energy conservation equations (\ref{eq:macrocons}) and (\ref{eq:microcons}), we obtain two coupled transport equations for macro and micro-processes in terms of the respective temperatures,  namely
\begin{equation}
 C_{M}\frac{dT_{M}}{dt}=\frac{\partial }{\partial x}\left(\lambda_{M}\frac{\partial T_{M}}{\partial x}+\lambda_{Mm}\frac{\partial T_{m}}{\partial x}\right),
 \label{eq:transportM}
\end{equation}
\begin{equation}
 C_{m}\frac{dT_{m}}{dt}=\frac{\partial }{\partial x}\left(\lambda_{mM}\frac{\partial T_{M}}{\partial x}+\lambda_{m}\frac{\partial T_{m}}{\partial x}\right),
 \label{eq:transportm}
\end{equation}
where the coefficients $\lambda_{ij}$ are thermal conductivities of direct and cross couplings. Note that even in the case that $\lambda_{Mm}=\lambda_{mM}$, the phenomenological coefficientes $L_{12}$ and $L_{21}$ are different.  

\section{Microstructure and time evolution}
\label{sec3}

Before we engange of the following section, we briefly discuss the transport and fitting parameters of the two temperature model. First, we have to differentiate between the $\lambda_{Mm}$ and $\lambda_{mM}$ coupling parameters and the coefficients $\lambda_{M}$ and $\lambda{m}$. The former coefficients should be considered part of the proposed model, but both $\lambda_{M}$ and $\lambda_{m}$ are transport properties of the material. $\lambda_{M}$ is the material bulk thermal conductivity and $\lambda_{m}$ contains the details of the microstructure of the system. Therefore, they can be determined only through experiments or by a more microscopic theory than the one proposed here.
This way, the two temperature model involves some fitting parameters, namely, the coupling parameters $\lambda_{Mm}$ and $\lambda_{mM}$, which will be taken to be equal and the heat capacities $C_{M}$ and $C_{m}$ which in the simulations will also be taken to be equal. In this way, the model has just two fitting parameters.

\subsection{Time evolution of the total temperature. Initial and boundary conditions}

Eqs. (\ref{eq:transportM}) and (\ref{eq:transportm}) will be solved by assuming that at one side of the system the measurable temperature is suddenly raised to $T_{h}$ while the opposite one is kept adiabatically isolated.  This implies Dirichlet-Neumann boundary conditions, that is 
\begin{equation}
 \frac{\partial T(0,t)}{\partial x}=0,
 \label{eq:leftbc}
\end{equation}

\begin{equation}
 T(L,t)=T_{h}-\Delta T_{h} Exp(-\frac{t}{t_{p}}),
 \label{eq:rightbc}
\end{equation}
The initial condition of the material is assumed to be a homogeneous distribution of the measurable temperature, i.e.,
\begin{equation}
 T(x,0)=T_{0},
 \label{eq:inicon}
\end{equation}
with $T_{0}$ being the room temperature. Then, the factor $\Delta T_{h}$ must be chosen equal to $T_{h}-T_{0}$. 

Eqs. (\ref{eq:leftbc}) and (\ref{eq:rightbc}) determine the boundary conditions for the macro and micro temperatures since they are related with the measurable temperature through Eq. (\ref{eq:total}). After that, Eq. (\ref{eq:leftbc}) implies that
\begin{equation}
 \frac{\partial T_{M}(0,t)}{\partial x}=0,
 \frac{\partial T_{m}(0,t)}{\partial x}=0,
 \label{eq:leftbcTMTm}
\end{equation}
and Eq. (\ref{eq:rightbc})
\begin{equation}
 T_{h}-\Delta T_{h} Exp(-\frac{t}{t_{p}})=\frac{C_{M}}{C_{M}+C{m}}T_{M}(L,t)+\frac{C_{m}}{C_{M}+C{m}}T_{m}(L,t).
 \label{eq:rightbcTMTm}
\end{equation}
Now it is reasonable to assume that the initial sudden increase of the measurable temperature at $x=L$ is reflected in the macro and micro temperatures in the same manner, that is to say,
\begin{equation}
 T_{M}(L,t)=T_{M0}-\Delta T_{M0} Exp(-\frac{t}{t_{p}}),
 \label{eq:pulsedbcTM}
\end{equation}
\begin{equation}
 T_{m}(L,t)=T_{m0}-\Delta T_{m0} Exp(-\frac{t}{t_{p}}).
 \label{eq:pulsedbcTm}
\end{equation}
\noindent which leads to the following conditions
\begin{equation}
 T_{h}=\frac{C_{M}}{C_{M}+C{m}}T_{M0}+\frac{C_{m}}{C_{M}+C{m}}T_{m0},
 \end{equation}
 \begin{equation}
 \Delta T_{h}=\frac{C_{M}}{C_{M}+C{m}}\Delta T_{M0}+\frac{C_{m}}{C_{M}+C{m}}\Delta T_{m0}.
\end{equation}

A few clarifying words on the previous assumption are necessary at this point. The boundary condition,  cf Eqs. (\ref{eq:pulsedbcTM}) and (\ref{eq:pulsedbcTm}), is established for the measurable -observable- temperature, defined in Eq. (\ref{eq:total}), which at time $t = 0$ rapidly increases to a $T_{h}$ value from a $T_{0}$ value. The question of how the sudden increase in measurable temperature translates to the macro and micro temperatures -which are not measurable- , has more
than one answer. We have given one that we think is physically reasonable. For example, assuming that only the macro temperature, or micro temperature, is affected when the measurable temperature increases at the boundary implies that the thermal excitation is transferred to only one of the subsystems. This is, of course, not reasonable. In fact, this is not the case even in ultra-fast phenomena where a system is pulsed with a laser, the duration of the pulse being on the order of picoseconds. In this example, although most of the injected energy is absorbed by the one of the subsystems, the
rest is absorbed by the other one. Later, through their interaction, some of the energy is also transferred to the second subsystem.

Now, it must be recalled that the micro temperature  describes the dissipative processes due to the presence of the internal structure of the system. As a consequence, it behaves like a fluctuating temperature around $T_{m0}$. In fact, the reference values $T_{M0}$ and $T_{m0}$ will be taken to be equal in the simulations of different structured solids made later on. This condition is a consequence of the fact that at $t=0$ the measurable temperature is homogeneous with value $T_{0}$. In this way, we have not only determined the boundary conditions for the macro and micro temperatures but it is reaffirmed the microtemperature as a thermodynamic variable and not as an internal variable. This point will be further considered in the discussion section.

Stationary profiles and the time evolution of  the macro and microscopic temperatures have been obtained by using the commercial Mathematica Wolfram software to solve Eqs. (\ref{eq:transportM}) and (\ref{eq:transportm}) (or (\ref{eq:nonltransportM}) and (\ref{eq:nonltransportm})) with conditions (\ref{eq:leftbcTMTm}), (\ref{eq:pulsedbcTM}) and (\ref{eq:pulsedbcTm}).  The results are presented in the following sections.

\subsection{On the internal structure of the solid}
\label{sec:internal}

\begin{figure}
 \centering
  \subfloat[Constant structure heat conductivity.]{
   \label{fig:micro1}
    \includegraphics[width=0.48\textwidth]{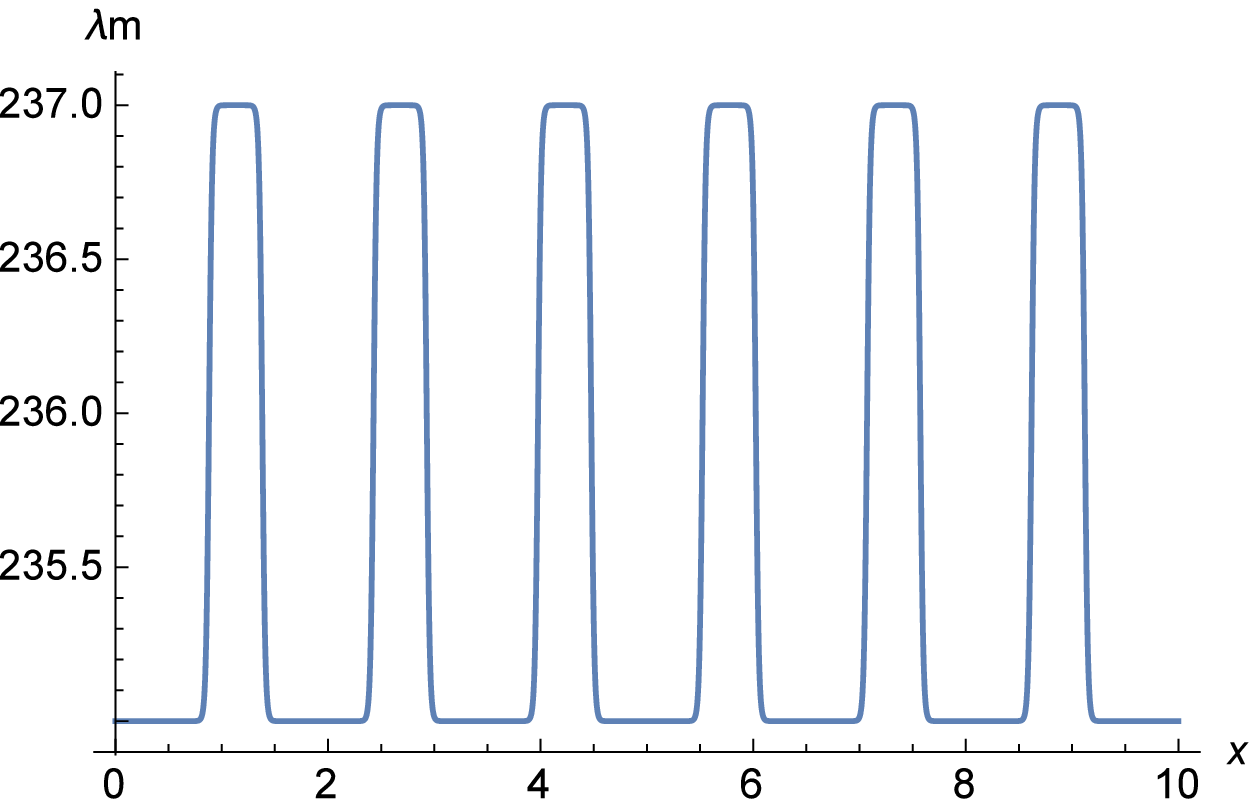}}
  \subfloat[Random structure heat conductivity]{
   \label{fig:micro2}
    \includegraphics[width=0.48\textwidth]{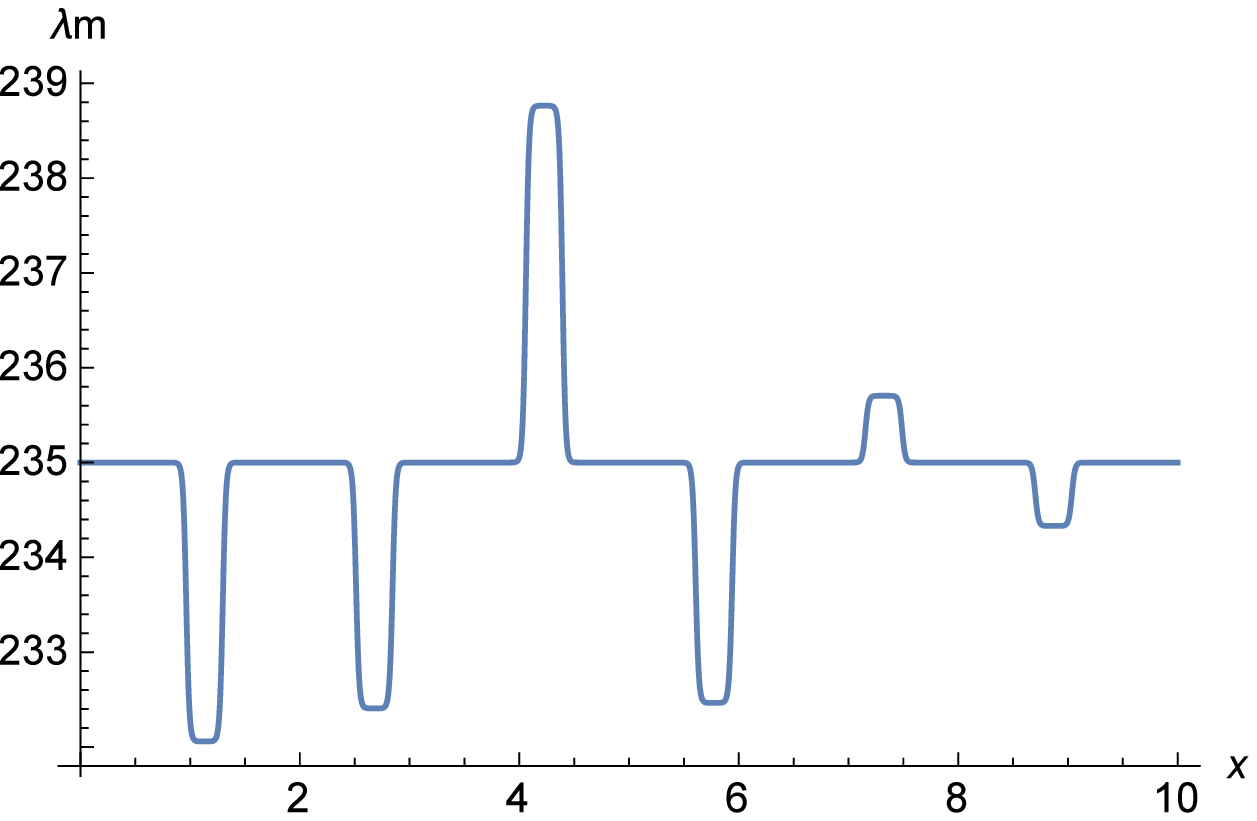}}
 \caption{Structure characterization through the heat microconductivity $\lambda_{m}$. The matrix heat conductivity (bulk) is 235 $W/mK$.}
 \label{fig:microstructure}
\end{figure}

In this short subsection we briefly describe in general terms the assumed internal structure of the studied solids. They are constituted by an alternated distribution of materials with different thermal conductivities. This distribution may be periodic or not. In this last case the materials may be randomly distributed in the space and even to have random values of their thermal conductivity. Figure \ref{fig:microstructure} shows generic examples of the structure introduced through $\lambda_{m}$ in the transport equations of the macro and the micro temperatures. In Figure \ref{fig:micro1} it can be seen a superlattice-like structure constituted by two materials with different constant thermal conductivities. In Figure \ref{fig:micro2} the same structure with random thermal conductivities of the second material is shown. This generic structures will be adapted to the specific characteristics of the studied systems.

\subsection{Near equilibrium processes}
\label{sec:OR}
\begin{figure}
\begin{center}
\includegraphics[scale=1.3]{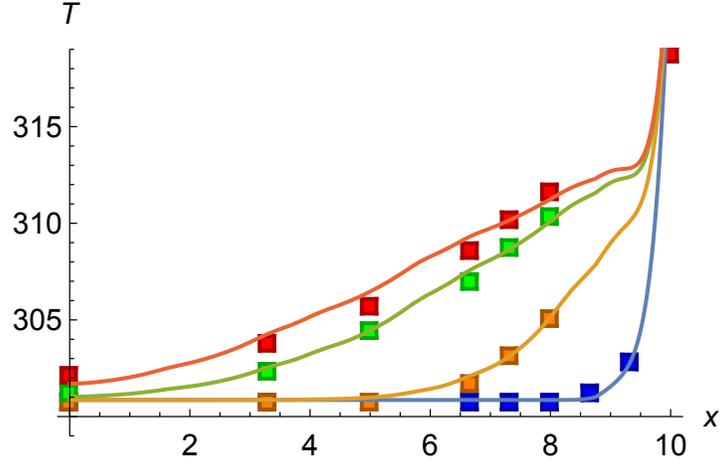}
\end{center}
\caption{\label{fig:cellular} Temperature profile (dots) in the cellular foam-like structure at different times from \"Ozdemir et al. \cite{ozdemir2008} and from the two temperature model (solid lines). Blue: 1 Sec., Yellow: 10 Sec. Green: 50 Sec., Red: 75 Sec. }
\end{figure}

\begin{figure}
\begin{center}
\includegraphics[scale=1.3]{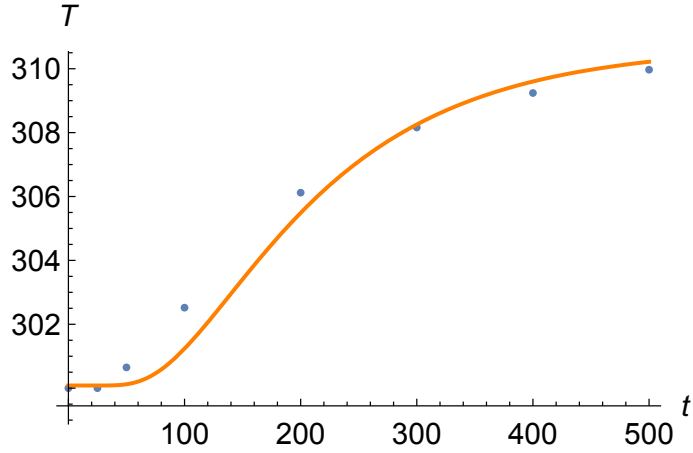}
\end{center}
\caption{\label{fig:cellular2} Time evolution of the temperature in the cellular foam-like structure at the non-exposed side. Dots represent the result from Lu et al. \cite{lu1999} and the solid line is the result from the two temperature model.}
\end{figure}

In this section we compare the prediction of the time evolution of the measurable temperature obtained from the two temperature model with that obtained by \"Ozdemir et al. \cite{ozdemir2008} through a homogenization technique. The system is a cellular foam-like structure subjected to Dirichlet-type boundary conditions.  In Figure \ref{fig:cellular}  the temperature profile for different times as obtained in Ref.  \cite{ozdemir2008} is shown.  We assume that inertial effects are not present (in agreement with the same  assumption already demonstrated in Ref. \cite{ozdemir2008}) and therefore that the local equilibrium hypothesis is valid along the process towards the stationary state. We use Eqs. (\ref{eq:nonltransportM}) and (\ref{eq:nonltransportm}) to obtain the temperature profiles in our structured material in order to qualitatively compare the two behaviors.  We have superimposed our results on  the graphs by \"Ozdemir et al \cite{ozdemir2008}. The fitting values of the coupling constants are in this case  $\lambda_{Mm}=\lambda_{mM}=0.1$ $W/mK$, $T_{M0}=310$ $K$, $T_{m0}=310$ $K$, $\Delta T_{M0}=\Delta T_{m0}=10$ $K$, and $t_{p}=10^{-3}$ $sec$.
As it may be observed, the temperature of the two systems shows a similar dependence on the position for the considered times.
Now we make a further comparison with the time evolution of the temperature in the non-exposed side of the same cellular structure obtained in \cite{lu1999} from a Fourier model with a boundary condition similar to that of Ref. \cite{ozdemir2008}. We have calculated  the time evolution of the temperature in the structure by using Eqs. (\ref{eq:nonltransportM}) and (\ref{eq:nonltransportm}). In Figure \ref{fig:cellular2} the response of the structure as a function of time from \cite{lu1999} is shown and we have also superimposed our result. The two curves are very close to each other. It is worth noting the behavior of the temperature for the earliest times. We note that the short time interval where the temperature remains constant is reproduced well by the two temperature model.

\subsection{Far from equilibrium processes}
\label{sec:far}
\begin{figure}
\begin{center}
\includegraphics[scale=0.73]{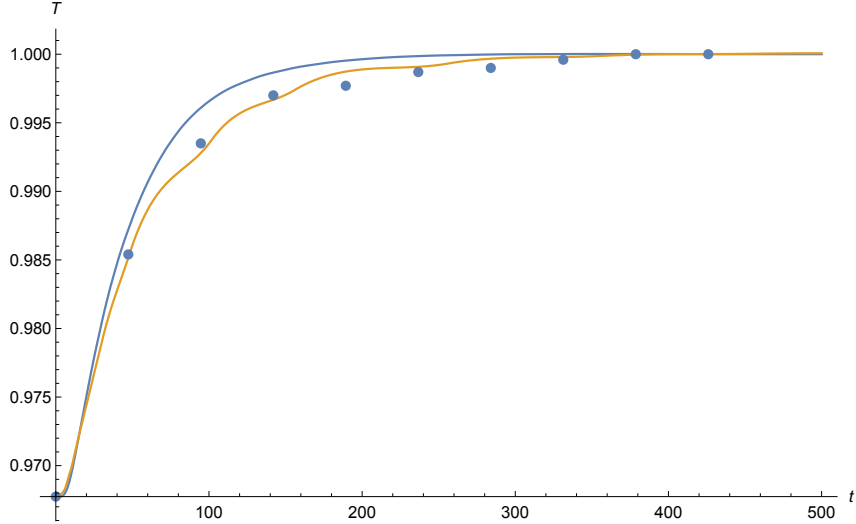}
\end{center}
\caption{\label{fig:history} Time evolution of the total temperature of pure Aluminum as derived from the Fourier model (solid blue line) and the proposed two temperature model (solid yellow line). Dots correspond to experimental data from Van et al. \cite{van2017}.}
\end{figure}

In this section we show that the two-temperature model proposed in this work, Eqs. (\ref{eq:transportM}) and (\ref{eq:transportm}), describes the non-Fourier regime found in heat transport experiments in structured solids at room temperature. Van et al. \cite{van2017} analyzed  four types of materials and measured the time thermal response of the material to a sudden increase of temperature at one side. We will focus here on the results for pure Aluminum. These results clearly show a non-Fourier behavior as it may be seen in Figure \ref{fig:history} where dots correspond to the experimental measurements and the solid blue line to the calculated time response in the Fourier regime obtained from

\begin{equation}
 \frac{\partial^2 T_{M}^{F}}{\partial x^2}=\alpha \frac{\partial T_{M}^{F}}{\partial t},
 \label{eq:fourier}
\end{equation}

\noindent with $\alpha$ the thermal diffusivity of bulk Aluminum. Note that we have obtained  the time evolution of the temperature using Eq. (\ref{eq:fourier}) subjected to the same initial and boundary conditions which are used to solve the two temperature model. 

At this point, one may reasonably wonder about the physical origin of the non-Fourier behavior given that the reported material is pure Aluminum. May one assume the existence of some kind of internal structure? The fact is that commercial pure Aluminum is subjected to the so called accumulative roll-bonding process at relatively high temperature which produces a lamellar structure characterizing the microstructure \cite{huang2003}. Rolled Aluminum then shows a structure similar to that assumed here for the solid and, as we stated in our initial assumptions, this microstructure is responsible for the non-Fourier behavior of the material. 

The solid yellow line in Figure \ref{fig:history} corresponds to the calculated thermal response of the material to the excitement as derived from the two temperature model,  Eqs. (\ref{eq:transportM}) and 
(\ref{eq:transportm}). The microstructure in this case is a resemblance of the inner structure of a superlattice: over a kind of matrix with the bulk heat conductivity of Aluminum, $237$ $W/mK$, a set of  slight variations of the heat conductivity (of the order of $2$ $W/mK$) is superimposed. An example is shown in Figure \ref{fig:microstructure}. The characteristic length of the microstructure is at least one order of magnitude smaller than the size of the system. The fitting values of the coupling constants are $\lambda_{Mm}=\lambda_{mM}=1$ $W/mK$ (see Eqs. (\ref{eq:transportM}) and (\ref{eq:transportm})).


The values $T_{M0}=310$ $K$, $T_{m0}=310$ $K$, $\Delta T_{M0}=\Delta T_{m0}=10$ $K$, and $t_{p}=10^{-3}$ $sec$ have been introduced in the boundary conditions,  Eqs. (\ref{eq:pulsedbcTM}) and  (\ref{eq:pulsedbcTm}).

As it can be seen, the solution obtained from Eqs. (\ref{eq:transportM}) and (\ref{eq:transportm}) agrees  well with the experimental points and   adequately describes the non-Fourier regime of heat transport found by Van et al. \cite{van2017}. The Fourier response (blue line) gives an idea of how much this regime deviates from the Fourier regime. We conclude that the non-Fourier regime described by the GK equation

\begin{equation}
 \tau\frac{\partial^2 T}{\partial t^2}-\frac{l^2}{\tau} \frac{\partial}{\partial t}\frac{\partial^2 T}{\partial x^2}+\frac{\partial T}{\partial t}-D \frac{\partial^2 T}{\partial x^2}=0.
 \label{eq:GK}
\end{equation}
where $\tau$, $l$ and $D$ stand for the relaxation time and mean free path of heat carriers and thermal diffusivity, respectively, is equivalent to the description based on the two coupled Fourier processes, Eqs. (\ref{eq:transportM}) and (\ref{eq:transportm}). This point will be discussed in more detail later in the final section.

\begin{figure}
\begin{center}
\includegraphics[scale=0.73]{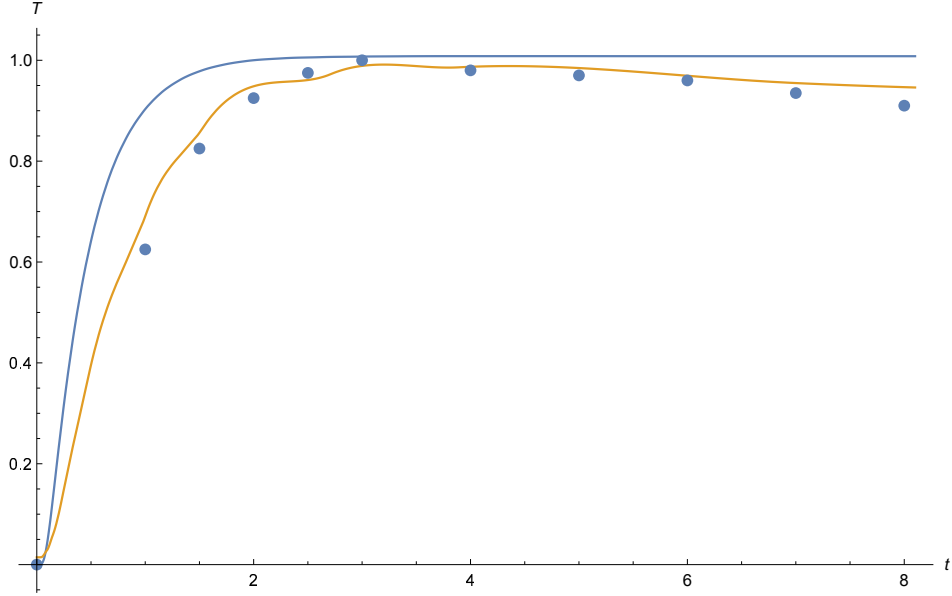}
\end{center}
\caption{\label{fig:history2} Time evolution of the total temperature of a biological tissue as derived from the Fourier model (solid blue line) and the proposed two temperature model (solid yellow line). Dots correspond to experimental data from Tang et al. \cite{tang2007}.}
\end{figure}

As a second example of structured materials we examine the results obtained by Tang et al. \cite{tang2007} for rapidly irradiated biological tissues. The reported behavior of the response of the system to the irradiation is different from that obtained for pure Aluminum. In this last case, the response does show an inertial effect since it lags behind the response of the Fourier regime. However, Figure \ref{fig:history} also indicates that there is no inertial effect in the approximation to the stationary state which occurs asymptotically from below the maximum temperature. In contradistinction, the response of the biological material shows not only inertia from the beginning but also when it approaches the stationary state from above the maximum temperature. Figure \ref{fig:history2} 
illustrates what has just been mentioned. Moreover, note that in the experiment the stationary temperature is smaller than the stationary temperature in the  Fourier regime. As before, the solution obtained from Eqs. (\ref{eq:transportM}) and (\ref{eq:transportm}) agrees  well with the experimental points and   adequately describes the non-Fourier regime of heat transport found by Tang et al. \cite{tang2007}. The fitting parameters used to obtain the calculated response, solid yellow line in Figure \ref{fig:history2}, are $\lambda_{Mm}=\lambda_{mM}=0.4$ $W/mK$ The biological tissue is constituted by meat with heat conductivity $0.47$ $W/mK$ and insertions of fat with $0.18$ $W/mK$. The boundary conditions are taken as in the previous case.  

Before ending this section, it is natural to wonder whether the Fourier heat transport model can reproduce the experimental data by simply incorporating the microstructure through the heat conductivity in Eq. (\ref{eq:fourier}). We have addressed this issue explicitly for Aluminum and the thermal response overlies on the solution of the Fourier process with constant heat conductivity (solid blue line) in Figure \ref{fig:history}. This reveals two important facts. On the one hand, that the non-Fourier behavior of the structured material comes from the influence of the micro temperature which contains the information of the microstructure. On the other hand, the effective or bulk part of the description of heat transport is blind to the details of the microstructure. These conclusions will be remarked in the final section.

\section{Microstructure and entropy production}
\label{sec4}
\subsection{Microstructure and heterogeneity}
In this section we study the relationship between microstructure and heterogeneity of the material. Heterogeneity is introduced in the spatial distribution of thermal conductivity by assuming that  it randomly varies along the material. To be more specific, we consider that the material is constituted by a material of $\lambda_{b}$ thermal conductivity (bulk) which changes in a random way in certain prescribed positions within the material. These variations are then included in $\lambda_{m}$, which is the heat conductivity associated with the microstructure.  
Now, we assume that the heat conductivity randomly changes its value at $N$ prescribed positions within the material. The random values are generated in the interval $(\lambda_{b}-2,\lambda_{b}+2)W/mK$. Let us denote the values generated in the indicated way as $\lambda_{i}$. We define the heterogeneity coefficient $C_{h}$ of the material as the mean squared deviation of the $N$ values in the microstructure from the bulk value of the heat conductivity, $\lambda_{b}$. That is, 

\begin{equation}
 C_{h}(N)=\sqrt{\frac{1}{N}\sum_{i=1}^{N}(\lambda_{i}-\lambda_{b})^2}.
 \label{coef}
\end{equation}

\subsection{Entropy production}

The local entropy production in the two temperature model is given by Eq. (\ref{eq:entropyprod}) which explicitly reads

\begin{eqnarray}
 \sigma\left(x,t\right)=\frac{1}{T_{M}\left(x,t\right)^2}\left[\lambda_{M}\left(\frac{\partial T_{M}}{\partial x}\right)^2+\lambda_{Mm}\frac{\partial T_{M}}{\partial x}\frac{\partial T_{m}}{\partial x}\right]
+\nonumber\\
\frac{1}{T_{m}\left(x,t\right)^2}\left[\lambda_{m}\left(x\right)\left(\frac{\partial T_{m}}{\partial x}\right)^2+\lambda_{mM}\frac{\partial T_{M}}{\partial x}\frac{\partial T_{m}}{\partial x}\right].
\label{eq:sigma2T}
 \end{eqnarray}

\begin{figure}
\begin{center}
\includegraphics[scale=0.98]{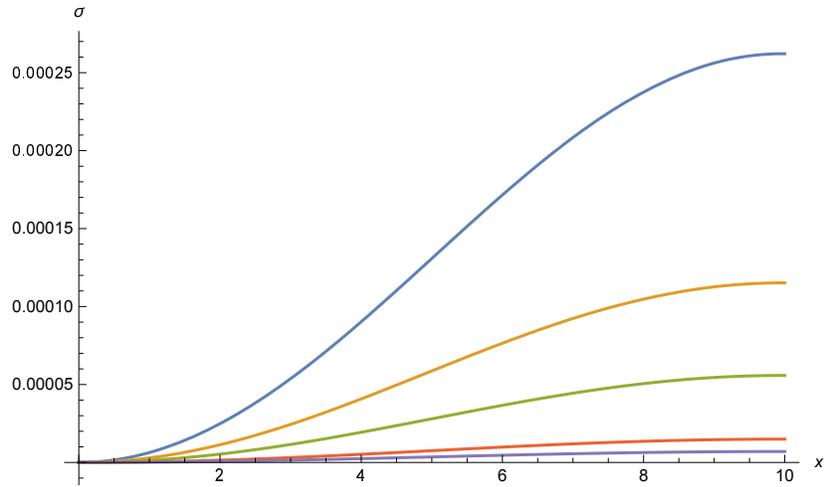}
\end{center}
\caption{\label{fig:Fig6} Entropy production profile at different times for pure Aluminum. Blue: 100, Orange: 120, Green: 150, Red: 180, Purple:220.}
\end{figure}

Later on,  Eq. (\ref{eq:sigma2T}) will be used to study the relationship between heterogeneity and entropy production in the stationary state. But before this, we refer to Figure \ref{fig:Fig6} where the profile of the  entropy production for different times in the case of pure Aluminum can be seen. We note that the region near the thermally excited boundary ($x=L$) is where more entropy is being produced. Also we observe that the entropy production for the early stages of the time evolution is greater than that for later times as the system approaches the stationary state. This suggests that the stationary state will obey a minimum entropy production principle, an issue that we will not further address here but that will be object of further research. 

At this stage, it is interesting to define an effective entropy production through the expression

\begin{equation}
 \sigma_{eff}\left(x,t\right)=-q\left(x,t\right)\frac{\partial}{\partial x}\left(\frac{1}{T}\right),
\end{equation}

\noindent where $T$ and $q$ stand for the total (measurable) temperature and the total heat flux, respectively. The last is given by

\begin{equation}
 q\left(x,t\right)=q_{M}\left(x,t\right)+q_{m}\left(x,t\right)
\end{equation}

\noindent with $q_{M}$ and $q_{m}$ given by the constitutive equations Eqs. (\ref{eq:constitutive1}) and (\ref{eq:constitutive2}). Therefore, 

\begin{eqnarray}
 \sigma_{eff}\left(x,t\right)=\left[\left(\lambda_{M}+\lambda_{Mm}\right)\frac{\partial T_{M}}{\partial x}+\left(\lambda_{Mm}+\lambda_{m}\left(x\right)\right)\frac{\partial T_{m}}{\partial x}\right]\times
 \nonumber\\
 \frac{\partial}{\partial x}\left(\frac{1}{aT_{M}+bT_{m}}\right),
 \label{eq:sigmaeff}
\end{eqnarray}

\noindent where $a=\frac{C_{M}}{C_{M}+C_{m}}$ and $b=\frac{C_{m}}{C_{M}+C{m}}$ (see Eq. (\ref{eq:total})).

\begin{figure}
\begin{center}
\includegraphics[scale=0.58]{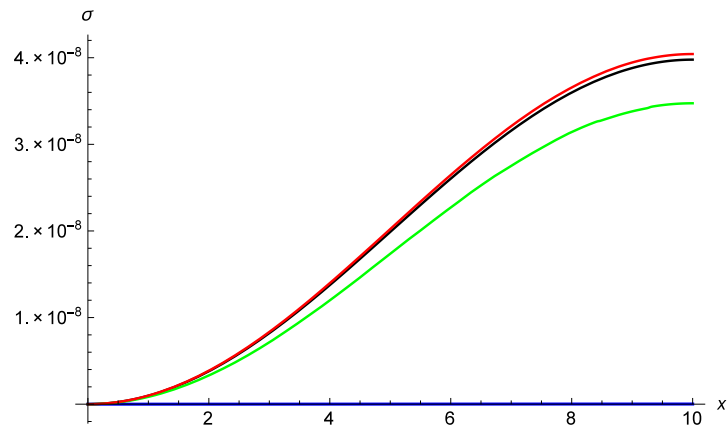}
\end{center}
\caption{\label{fig:Fig1} Stationary state entropy production profile  for pure Aluminum. Blue: Fourier,  Red: two temperature model (Eq. \ref{eq:sigma2T}),Green: effective (Eq. \ref{eq:sigmaeff}), Black: total (measurable) temperature (Eq. \ref{eq:sigmacll}).}
\end{figure}

A comparison of the effective entropy production and that coming from the two temperature model in the stationary state can be seen in Figure \ref{fig:Fig1}, where they are represented by the green solid line and the red solid line, respectively. Clearly, the entropy production in the effective medium is smaller than that of the solid with internal structure. In Figure \ref{fig:Fig1} we have included the entropy production in the Fourier process (blue solid line) and the entropy production given by the classical expression \cite{deg1984}

\begin{equation}
 \sigma_{cl}\left(x,t\right)=\frac{\nabla T^2}{T^2},
 \label{eq:sigmacll}
\end{equation}

\noindent represented by the black solid line. As expected, the entropy production in the Fourier process is the smallest (the Fourier solid does not consider the internal structure at all) and those from the two temperature model and the classical expression coincide. It is then found that the microstructure introduces a new dissipation mechanism in the system. It must be considered the fact that  the microstructure involves the presence of interfaces where entropy is produced but not heat, as it occurs in
the cases we have studied. In fact, at no point in the material there is an increase in temperature due to heating, which would imply the presence of a source term in the energy equation. By dissipation we are strictly understanding the processes associated with entropy production.\\

In the following section the heterogeneity coefficient as obtained above is used to examine the relation between heterogeneity and stationary entropy production within the two temperature model.

\subsection{Heterogeneity and entropy production}
\begin{figure}
\begin{center}
\includegraphics[scale=0.73]{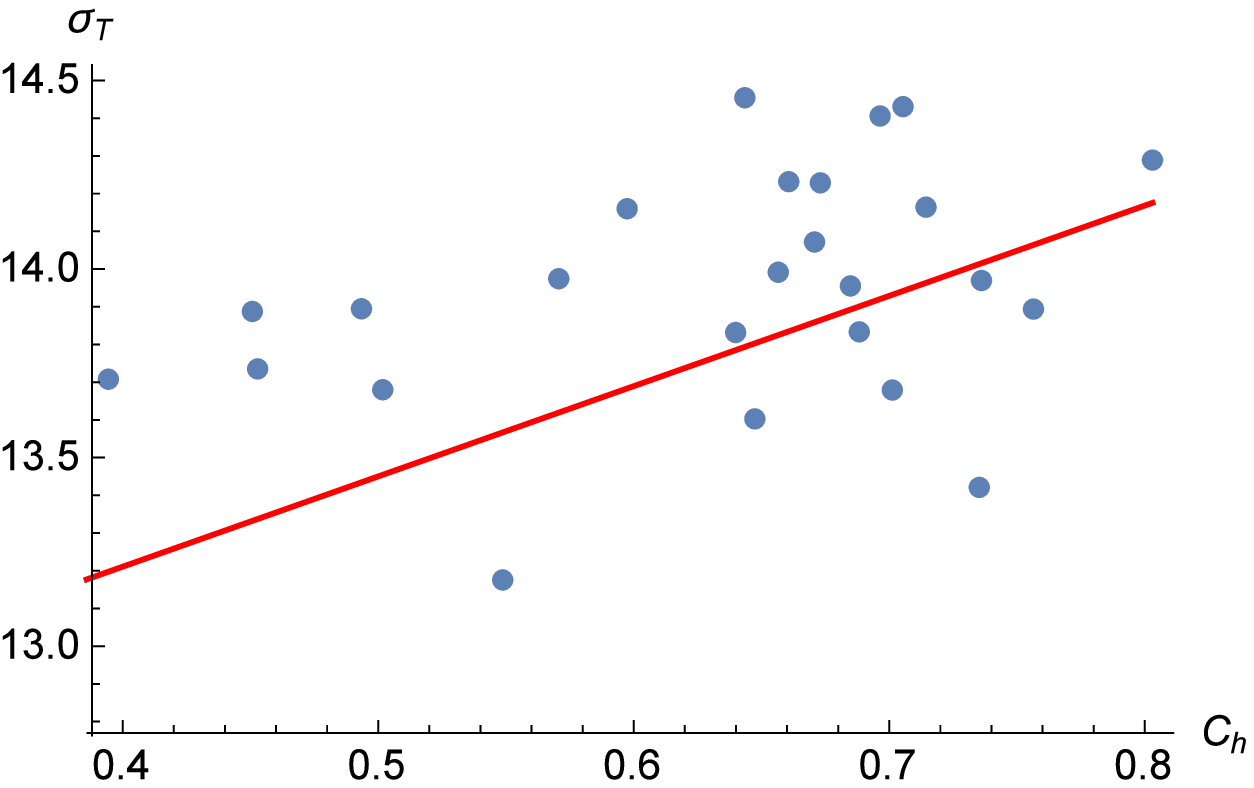}
\end{center}
\caption{\label{fig:Fig7a} Total entropy production vs. heterogeneity coefficient for pure Aluminum. Case: increasing random thermal conductivity as a function of position. The values of the integrated entropy production are affected by a factor $10^8$. $\sigma_{T}=12.25+2.39\,C_{h}$.}
\end{figure}

\begin{figure}
\begin{center}
\includegraphics[scale=0.73]{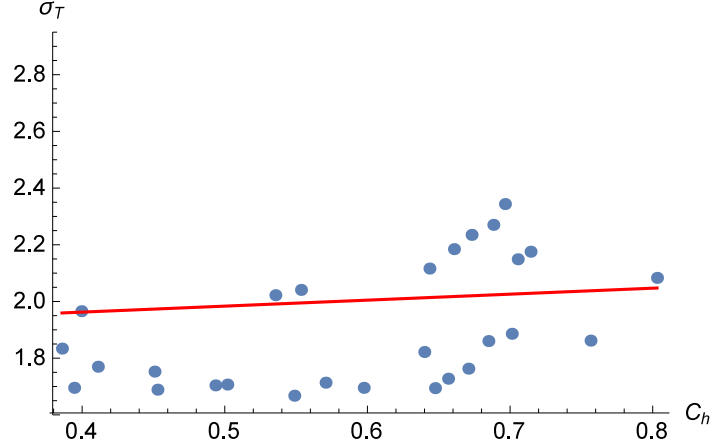}
\end{center}
\caption{\label{fig:Fig7b} Total entropy production vs. heterogeneity coefficient for pure Aluminum. Case: decreasing random thermal conductivity as a function of position. The values of the integrated entropy production are affected by a factor $10^8$. $\sigma_{T}=1.88+0.21\,C_{h}$.}
\end{figure}

 We now wonder whether a relationship exists between heterogeneity and the total entropy production (summed over the entire spatial domain). We may only answer this question to a limited extent due to the randomness introduced by the fluctuating nature of the heat conductivity at the microscale. We restrict ourselves to the case where the random values of the heat conductivity at the microscale show an increasing or decreasing space profile. We then consider several realizations of the conductivity profile and calculate the total entropy production for each one of them. Finally, a linear regression analysis to the data is performed. The result may be seen in Figures \ref{fig:Fig7a} and \ref{fig:Fig7b} for the increasing and the decreasing cases, respectively. Both show that there exists a positive correlation between heterogeneity as measured by the heterogeneity coefficient and the value of the entropy production.

\section{Discussion and concluding remarks}
\label{sec5}

\subsection{Comparison with other theories}

The two temperature model proposed in this work has some similarities and some  differences with several physical theories aimed at describing the influence of the structure of solids on their thermal behavior. The first part of this final section will be devoted to expose similarities and differences with some of them related through their main assumptions although their subsequent procedures may be quite different. Then we will discuss our results and make some concluding remarks. 

\subsubsection{Multifield theory}

This theory, which in fact encompasses several other theories, considers the system as constituted by material patches \cite{mariano2002}, which are in fact volume elements characterized by their position in the system, and an order parameter which contains the information on the substructural configuration of each patch. The order parameter is chosen such that it is an observable quantity and its changes are considered to be responsible for non-Fourier behaviors in heat conduction provided they are sensitive to temperature variations \cite{mariano2017}. In this formalism, like in ours, Fourier 's law is not changed. Unlike ours, the effects of the change of microscopic mechanisms are considered in the energy conservation equation and  a sort of constitutive relationship between the geometric descriptors of the microstructure and the temperature is introduced. 

\subsubsection{Homogenization techniques}

The homogenization process, used in \cite{ozdemir2008} to describe the thermal behavior of the Aluminum structure studied in Section \ref{sec:OR}, presents some similarities with the two temperature scheme. It is based on the existence of a macro and a micro length scales in the system and respective thermal processes. The temperature at the macro level evolves in time according to energy conservation and at the micro level it is assumed that the energy storage in the representative volume element vanishes, which results in a stationary state. There is not explicit coupling of both processes until the homogenization process is implemented. Then both temperature fields are obtained through standard multi-scale analysis.

\subsubsection{Two temperature models}

It is well known that in solid rigid conductors heat can be transported by one or more types of carriers \cite{ma2013,sobolev2016,chen2002,kovacs2018,vazquez2020}. Experiments in superlattices have made the wave properties of phonons evident and their effects have been interpreted within the coherent-incoherent phonon scheme \cite{luckyanova2012,ravichandran2013,saha2016,saha2017}. All the mentioned examples have in common representations in terms of two (or more) temperatures which are widely used in the study of heat transport problems of current scientific and applied interest. Formally, they have been obtained within classical irreversible thermodynamics and in some cases within extended irreversible thermodynamics \cite{sellito2016}. 
In this subsection we make some further remarks on the first main result of this work, namely, the relationship of the two temperature model of Section \ref{sec:thermo} and the GK equation.  

Splitting the heat flux into two parts or introducing an additional temperature in the GK equation lead to a simplification of the heat transport model. In the first case, the description is reduced (separated) to a Fourier process and a Maxwell-Cattaneo process for each of the parts of the splitted heat flux, respectively \cite{van2012,ciancio2016}. In the second case, the description is reduced to a Fourier process and a heat exchange equation \cite{van2012}. Similarly, the two temperature model of this work shows that the non-Fourier regime of heat transport in structured materials is the result of the coupling of two Fourier processes which take place at the macroscopic and the microscopic length scales, respectively. The solutions of Eq. (\ref{eq:GKh}) and Eqs. (\ref{eq:transportM}) and (\ref{eq:transportm}) are coincident and they describe well the experiments concerning heat transport in the structured material as shown in Figure \ref{fig:cellular}.

The GK equation, Eq. (\ref{eq:GK}), contains a hierarchy of Fourier processes as it was suggested by Van et al. \cite{van2017} and also earlier by others \cite{berezovski2011,engelbrecht2015}. Such hierarchy may be made apparent if the equation is rewritten in the form

\begin{equation}
 \tau\frac{\partial}{\partial t}\left(\frac{\partial T}{\partial t}-\frac{l^2}{\tau} \frac{\partial^2 T}{\partial x^2}\right)+\left(\frac{\partial T}{\partial t}-D \frac{\partial^2 T}{\partial x^2}\right)=0.
 \label{eq:GKh}
\end{equation}

In this way, the non-Fourier regime being described by Eq. (\ref{eq:GKh}) is the result of a Fourier process (second parenthesis in Eq. \ref{eq:GKh}) and the time evolution of a second Fourier process represented in the first parenthesis of the same equation. It must be recalled that $T$ is the measurable temperature given by Eq. (\ref{eq:total}). This interpretation of the GK equation retains some differences with the addition of physical fields to the description. Firstly, because there is only one temperature field and secondly, because there are no separate time evolution equations for each of the Fourier models forming the hierarchy. Nevertheles, it is interesting to note that the specific realization of the first of the Fourier process is determined by the diffusivity of the system and the second one by the phonon mean free path and the relaxation time. It is also worth mentioning that the mean free path-relaxation time factor characterizes the deviation of the process from Fourier behavior. This indicates that the  non-Fourier process results from two Fourier processes occurring at different length scales.  This is reminiscent of the main assumption made here to build the two temperature model.

\subsubsection{Internal variables thermodynamic theory}
Though we named $T_{m}$ as the microtemperature, our concept maintains some differences with the microtemperature from the internal variables theory \cite{berezovski2016}. First, the microtemperature in this framework like ours, refers to those processes which occur at a microscopic length scale emerging from the internal structure of the system and influencing the behavior of the measurable temperature. However, in contradistiction with the internal variables approach, the microtemperature in this work is not only assumed, but in fact also shown, to be determined by the conditions at the boundaries of the system.

As mentioned before, the microtemperature introduces a non-Fourier effect on the time evolution of the measurable temperature which is a resemblance of the non locality in the time (memory) introduced by the time derivatives in the constitutive equations. This was named by Van \cite{van2003} as the relocalization effect. In this way, although the transport equation (in terms of the measurable temperature) remains being of the parabolic type, the  effect of the microtemperature is equivalent to the presence of the time derivatives of the heat fluxes in the constitutive equations, (Eqs. \ref{eq:constitutive1}) and (\ref{eq:constitutive2}). In the context of the internal variables theory,  it is necessary to consider an extra internal variable and getting the  dual  internal variable scheme to obtain the relocalization effect as it was formally shown in \cite{berezovski2016}. The introduction of the second variable leads to constitutive equations of the Cattaneo type and a hyperbolic transport equation.

\subsection{Further concluding remarks}

We summarize in this last section the main conclusions drawn throughout the manuscript.

The comparison of the two temperature model for structured solids with experimental results has shown that the model reproduces well the non-Fourier behavior in two cases, namely, when the system is externally excited in a short time scale in such a way that it transits through non-equilibrium states which do not satisfy the Onsager reciprocity relations and also when the system is slowly excited in such a way that it transits through states satisfying the Onsager relations.

The results obtained reveal two important facts. On the one hand, that the non-Fourier behavior of the structured material comes from the influence of the microtemperature which contains the information of the microstructure. On the other hand, it leads to the conclusion that the Fourier description of heat transport is blind to the details of the microstructure.

The introduction of the microtemperature in the description seems to incorporate the important property of the inertia in the time evolution of the system. This is equivalent to the relocalization concept established by Van \cite{van2003} and described by other non-Fourier models like those of Cattaneo and Guyer-Krumhansl.

It must be stressed that the microtemperature of the two temperature model is not comparable to the microtemperature of the internal variables theory, since the former is determined through  the boundary conditions on physical grounds, see Eq. (\ref{eq:pulsedbcTm}). Nevertheless, the concept of microtemperature in this work and the procedures that we followed may help to clarify the physical nature of the microtemperature in the internal variables theory \cite{berezovski2016}. This point certainly deserves a more profound examination. Finally, we want to mention that perhaps the closest formalism to ours is that of internal variables. It is well known, however, that the introduction of internal variables makes it possible to describe a system with an internal structure without making any assumptions about the structure itself \cite{berezovski2018waves}. In this aspect, the scheme maintains a great difference with the two-temperature scheme since it is necessary to include the details of the structure to describe the behavior of the measurable temperature. However, they bear a great similarity in the sense that the introduction of the internal variables or the two temperatures allows to describe memory and non-local effects (this was not formally shown in the two-temperature scheme)

We have shown, to a limited extent, that entropy production and heterogeneity are correlated. In the case of materials with random increasing heat conductivity, the more heterogeneity the more entropy production observed in the system. We have two additional comments in this respect. On the one hand, the time behavior of the entropy production shown in Figure \ref{fig:Fig6} suggests that the stationary state obeys an extremum variational principle. This will be further investigated in future work. On the other hand, as it may be seen in Figure \ref{fig:Fig1}, the presence of the internal structure in the system produces a further dissipation mechanism causing a bigger entropy production as compared with the Fourier solid and the effective medium. In this context, the fact that the entropy production from the two temperature model, (Eq.\ref{eq:sigma2T}), coincides with that calculated from the classical expression, Eq.(\ref{eq:sigmacll}) , by using the total measurable temperature, Eq. (\ref{eq:total}) is also remarkable. 

We have left at the very end a comment on the physical nature of the non-Fourier heat transport regime experimentally observed in solids at room temperature. Our results suggest that the transport mechanism in such kind of phenomenon may be different from that of hyperbolic heat conduction. In spite of the good agreement of the predictions based on GK equation with the experimental results, as it was shown in \cite{van2017}, the understanding of the non-Fourier transport may not require the introduction of second sound and wave behavior considerations. The GK equation predicts those behaviors at short time scales of the order of the relaxation time of the heat flux. It seems likely those rapid phenomena may not be determining the non-Fourier heat transport regime observed in solids at room temperature.  

\subsection*{Acknowledgement}
FV acknowledges fruitfull discussions with Peter Ván (Budapest), Robert Kovács (Budapest) and Sergey Sobolev (Moscow) on related topics.

\subsection*{Funding}
  This work was  supported by PRODEP (México), IER-UNAM and CONACYT (México) under grant 258623.

\bibliographystyle{unsrt}  


\end{document}